# Growth mechanism of cluster-assembled surfaces: from sub-monolayer to thin film regime

Francesca Borghi[1], Alessandro Podestà[1,*], Claudio Piazzoni[1], Paolo Milani[1,*]

[1]Centro Interdisciplinare Materiali e Interfacce Nanostrutturati (C.I.Ma.I.Na.) and Dipartimento di Fisica, Università degli Studi di Milano, via Celoria 16, 20133 Milano, Italy.

[*] Corresponding authors addresses: alessandro.podesta@mi.infn.it; pmilani@mi.infn.it

**Abstract**

Nanostructured films obtained by the assembling of preformed atomic clusters are of strategic importance for a wide variety of applications. The deposition of clusters produced in the gas phase onto a substrate offers the possibility to control and engineer the structural and functional properties of the cluster-assembled films. To date the microscopic mechanisms underlying the growth and structuring of cluster-assembled films are poorly understood, and in particular the transition from the sub-monolayer to the thin film regime is experimentally unexplored. Here we report the systematic characterization by Atomic Force Microscopy of the evolution of the structural properties of cluster-assembled films deposited by Supersonic Cluster Beam Deposition. As a paradigm of nanostructured systems, we have focused our attention on cluster-assembled zirconia films, investigating the influence of the building blocks dimensions on the growth mechanisms and on the roughening of the thin films, following the growth process from the early stages of the sub-monolayer to the thin film regime. Our results demonstrate that the growth dynamics in the sub-monolayer regime determines different morphological properties of the cluster-assembled thin film. The evolution of roughness with the number of deposited clusters reproduces exactly the growth exponent of the ballistic deposition in the 2+1 model, from the sub-monolayer to the thin film regime.

# I. INTRODUCTION

The use of thin films is ubiquitous in a very large number of applications ranging from microelectronics and photonics to energy conversion and biomedicine, where an upmost coating with thickness of only a few tens, or hundreds of nanometers provides improved functionalities to a bulky device[1].

Thin film fabrication technologies are based on the precise assembling of atoms and molecules as elemental building blocks[1], as, for example, in molecular beam epitaxy and atomic layer deposition[1,2]. In general, physical and chemical vapour deposition methods consist on the assembling of precursors from the gas phase onto a substrate, where nucleation and growth of thin films is determined by adsorption, surface diffusion, chemical and physical binding[1].

The presence of nanoscale structures or defects in thin films has important consequences on their structural and functional properties: the decrease in electrical conductivity compared to the bulk in polycrystalline thin metal films[3,4], the variation in magnetic properties[5] as a function of film thickness and growth conditions. The precise understanding and control of the presence of defects in thin films offer the opportunity to fabricate structures with novel structural and functional properties thus turning a vice into a virtue[6].

Atomic clusters consisting of aggregates from few to several thousand atoms, have been proposed as building blocks of nanostructured solid state systems and devices with unique structural, electronic, optical, magnetic, and catalytic properties[7,8]. Following the systematic study of clusters in the gas phase[9], several groups focused their attention on the use of free clusters for the assembling of nanostructured thin films[10–14]. Many scholars, in particular theorists, suggested that size-selected clusters could be assembled as "super-atoms" to form ordered crystalline structures in analogy with the approaches developed for atom-assembled films[15]. To date the only systems assembled in macroscopic quantities from size-selected clusters in the gas phase are those fabricated using fullerenes[16]. As an alternative to the use of size-selected aggregates, clusters with a broad mass distribution has been recognized, in the last decade, as candidates to assemble, on a large scale, systems with very interesting nano- and mesoscale properties[17–19].

Cluster Beam Deposition (CBD)[10,8-20] is a technology for the fabrication of nanostructured thin films and devices, since it allows the deposition on a substrate of neutral and ionized nanoparticles produced in the gas phase[21]. CBD has been proven to be a powerful bottom-up approach for the engineering of nanostructured thin films with tailored properties, resulting from the so-called 'memory effect', i.e. the fact that the nanoscale building blocks maintain their individuality during the assembling process[10,14,20]. Among different approaches to CBD, Supersonic Cluster Beam Deposition (SCBD)[21,22] presents several advantages in terms of deposition rate, lateral resolution (compatible with planar microfabrication technologies) and neutral particle mass selection process by exploiting aerodynamic focusing effects[22–25]. All these features make SCBD a very effective tool to fabricate nanostructured films with novel structural and functional properties[19].

SCBD can be used to produce surfaces with multi-scale controlled disorder as substrates to study quantitatively the effect of nanoscale topography on biological entities[26]. In fact, this method is able to sufficiently explore the parameter space of topographical cues by the rapid parallel fabrication of surfaces with different nanoscale topographies and subsequent high-throughput assaying of these surfaces exposed to the different conditions that can affect their biological activity[19,27]. To this purpose, transition metal oxide clusters, titania and zirconia in particular, have been used[19,27,28]. We have recently demonstrated that cluster-assembled zirconia surfaces represent an ideal playground to study the interactions of nanostructured interfaces with biological entities, as for example the modulation of the cellular biological functions[19,29,30].

The assembling of clusters by SCBD produces nanostructured films with a nanoscale topography whose roughness can be accurately controlled and varied[27]: the morphology of cluster-assembled materials is characterized by a hierarchical arrangements of small units in larger features up to a certain critical length-scale, determined by the time of the deposition process[31]. Cluster-assembled film are characterized by high specific area and porosity at the nano and sub-nanometer scale, extending in the bulk of the film[32-33]. The control and manipulation of these structural properties offer the possibility to fabricate nanostructured systems with tailored properties in an efficient and scalable way. It is thus very important to understand the basic mechanisms of the early stages of cluster assembling on surfaces in order to identify the ingredients for the control and engineering of larger nanostructures.

Providing theoretical models to describe the sub-monolayer growth of cluster-assembled films is challenging. Tentatively, one could consider each cluster as a 'super atom', and extrapolate the predictions of theoretical studies, originally developed for the atomic deposition. However, by comparing the deposition of preformed clusters to the atomic deposition, many differences become evident, due to the inner structure of the clusters: the possibility of two clusters merging into a larger cluster[34]; different mechanisms of clusters diffusion[10,34]. According to the percolation model,[35,36] particles do not diffuse after being deposited; this theory forbids therefore the aggregation of the diffusing particle. Other models describe diffusing particles that aggregate, such as the Cluster-Cluster Aggregation (CCA) model;[37] these models do not allow however for the continuous injection of new particles via deposition. Neither the percolation nor the CCA models do strictly apply to our case. A model that incorporates the three main physical mechanisms of the clusters-assembled thin film growth is the Deposition, Diffusion and Aggregation (DDA) model,[38] which introduces the possibility of cluster diffusion, although it is based on some limiting assumptions, such as the simple juxtaposition of two separated entities, and not their coalescence into a new larger one. This model is very useful to describe the evolution of the islands morphology and density on the substrate, in the case of a constant flux of particles. The DDA model describes the evolution of fractal structures, which are also characteristic of the models cited above. The situation is further complicated if one considers that the injection of particles, as in the case of SCBD, is pulsed, and there is a broad, and possibly multimodal distribution of clusters sizes[27].

Here we report the results of a systematic characterization by Atomic Force Microscopy (AFM) of the evolution of morphological properties of nanostructured Zirconia (ns-$ZrO_2$) cluster-assembled films, deposited by SCBD, from the sub-monolayer to the thin film regime. In particular, we have investigated the influence of the size of the building blocks on the growth mechanisms and on the final surface morphology of nanostructured films.

Due to the complexity of the system under study (clusters size dispersion[27], different diffusivity of clusters depending on their dimension, pulsed deposition regime[39], possible coalescence phenomena), and to the inaccuracy of the assessment of the lateral dimensions of particles due to AFM imaging limits[40], the quantitative description of the lateral (x-y) growth of clusters/islands as described in terms of classical models turns out to be very difficult. We have therefore focused our attention on the quantitative description of the evolution of the vertical width (the rms roughness) of the interface, and its scaling[41].

## II. EXPERIMENTAL METHODS

### A. Production and deposition of $ZrO_x$ clusters

Zirconia clusters have been produced and deposited with a SCBD apparatus equipped with a Pulsed Microplasma Cluster Source (PMCS)[39]. Details on this technique have been extensively presented elsewhere[14,21,22], here we discuss only those aspects relevant for the sub-monolayer deposition.

The apparatus consists of two differentially pumped vacuum stages. A PMCS is mounted outside the first chamber (expansion chamber) on the axis of the apparatus. The PMCS is operated in a pulsed regime: a solenoid pulsed valve, facing one side of the source cavity injects high-pressure inert gas (He or Ar) pulses with duration of few hundreds of microseconds at a repetition rate of 4 Hz. The gas injection is followed by a very short (a few tens of μs) and intense (a few hundreds of amperes) electrical discharge between the cathode (zirconium rod) and an anode buried in the source body. Due to aerodynamic effects, a localized high pressure region is formed at the cathode target surface and ablation of the metallic target through ion bombardment is thus confined; subsequent condensation of sputtered atoms results in cluster nucleation[42]. The inert gas-clusters mixture is then extracted from the PMCS into high vacuum (p ~$10^{-6}$ mbar) through a nozzle, and it expands to form a seeded supersonic cluster beam. The nozzle is connected with a series of aerodynamic lenses used to focus neutral nanoparticles on the beam axis[23,43].

Cluster deposition takes place in the deposition chamber where the supersonic beam impinges on substrates mounted on a x-y-z motorized sample holder. Fig. 1 shows a schematic representation of the deposition process on the sample holder hosting several substrates of polished silicon intercepting different portions of the cluster beam with an approximately Gaussian intensity profile[22]. Si substrates (1x0.5 $cm^2$ with a RMS roughness is 0.08 ± 0.01 nm) are typically cleaned in acqua regia, ethanol, and then dried in a nitrogen flux before deposition.

The substrates holder can rotate around its vertical axis in order to intercept the beam for a controlled amount of time or number of shots: a single rotation period correspond to the deposition of four pulses from the PMCS. The clusters deposited from this number of pulses maintain on the substrate a mass distribution very similar, if not identical, to that of the free clusters produced in the PMCS and carried by the supersonic expansion. By increasing the deposition time (number of pulses) the morphology of the objects deposited on the substrates changes and evolves in *island*s with a structure resulting from coalescence or juxtaposition of the primeval clusters.

In the PMCS we produce prevalently metallic Zr clusters, due to the presence of small traces of oxygen in the stagnation cavity (the purity level of the gas is N6.0=99.9999%)[44]. The clusters are deposited in a deposition chamber characterized by a pressure of $10^{-6}$ mbar. So a substantial oxidation of the clusters takes place very rapidly[45] because of the interaction of the Zr clusters with free oxygen and water molecules[44], and later on, upon exposure of the sample to air, resulting in cluster-assembled nanostructured $ZrO_x$ films (ns-$ZrO_x$) with x close to 2.[28] Clusters have a broad mass distribution that has been characterized by AFM after deposition (vide infra). The analysis of the particle size distribution, deposited with Argon as carrier gas, from TEM images[28] provides an average clusters diameter of 6.0 ± 1.7 nm.

The cluster beam profile is approximately Gaussian[22]: the largest particles are concentrated along the beam axis and the cluster diameter decreases going from the beam center to the periphery[46]. The mass distributions of the deposited clusters depend on the carrier gas (Helium or Argon), here we present the results obtained by depositing clusters by using both inert gases; the obtained samples are named ns-$ZrO_x$/He and ns-$ZrO_x$/Ar, respectively.

We characterized the evolution of the samples morphological properties as a function of the surface coverage[34], which is defined as the ratio between the projected area occupied by clusters on the surface and the scanned area and which increases with the number of multiple shots.

We have analyzed the cluster-assembled films with increasing values of coverage (Table I) and with different cluster size distributions depending on the regions of the beam selected for the deposition.

**TABLE I**. The ns-ZrO$_x$ samples analyzed in this study.

|  | ns-ZrOx/He Center of the beam | ns-ZrOx/He Periphery of the beam | ns-ZrOx/Ar Center of the beam | ns-ZrOx/Ar Periphery of the beam |
|---|---|---|---|---|
| # of samples analyzed | 7 | 7 | 15 | 15 |
| Range of coverage (%) | 1 - 95 | 2 - 87 | 3 - 98 | 4 - 88 |

It is important to notice that the incident flux $f$ of clusters on the substrate during the deposition is quite different for the systems analyzed ($f \approx 1 \cdot 10^{10}$ clusters/s·cm$^2$ for ns-ZrOx/Ar and $f \approx 1 \cdot 10^{11}$ clusters/s·cm$^2$ for ns-ZrOx/He), as inferred from our data, based on the measured number of particles, total area investigated, and deposition time. The deposition rate in terms of mass per unit time, however, is different, since the average size of deposited particles is different.

**B. Atomic Force Microscopy characterization**

For each sample, different images (typically nine) with a scan area of 2μm x 1μm and a sampling frequency of 1nm/pixel e 2nm/pixel in x and y directions, respectively, have been acquired using a Multimode Nanoscope IV AFM (Bruker). The AFM was operated in tapping mode in air with scan rates of 1 Hz and small free oscillation amplitudes (10nm). Silicon tips with radius below 10nm and resonance frequency of 300 kHz have been used.

AFM images have been prepared for the analysis by subtraction of 2$^{nd}$-order polynomials, line by line, in order to remove the tubular scanner bow and the tilt of the sample, and later by applying a median filter with a 3x3 kernel. A mask has been built for each image in order to identify the objects of interest on the surface. To this purpose a z-threshold value was set at two standard deviations above the mean value of the background (2σ ~ 0.2 nm). Fig. 2a shows a typical AFM top-view map of sub-monolayer ns-ZrOx/Ar morphology, while Fig. 2b shows the three-dimensional topographic map (bottom) and its corresponding mask (top) face to face, in order to highlight the selection process.

Each AFM image contains multiple objects of interest. For each object, we identify the following morphological properties:

- Height (nm), as the difference between the median value of the five highest points of the object and the mean value of the background;
- (projected) Area ($nm^2$), as the sum of the area of all pixels in the object;
- Volume ($nm^3$), is the result of the numerical integration of the height profile over the area occupied by the object.

Since the measurement of heights by AFM is not affected by the tip-sample convolution effect[47], and assuming that the primeval clusters are spheroidal objects, we take the height as an effective measure of the particle diameter. Because of the mentioned tip-sample convolution effect, the evolution of the different objects in x-y directions can be characterized only qualitatively[48].

The samples with the lowest coverage (four pulses) have been analyzed to characterize the size distribution of the primeval incident clusters. To this purpose only globular objects have been selected for the analysis, by applying the following selection criteria: a linear relationship in semi-log scale between volume and height or between equivalent radius and height; axes ratio in the range between 0.6 and 1; height below 20 nm. The identification of the globular objects with the primeval clusters is based on the following assumptions:

- diffusion-induced juxtaposition or coalescence phenomena lead to lateral growth of the primeval clusters, and to a deviation from the spheroidal geometry;
- the flux of incident particles is such that the mean distance between deposited clusters is large enough to make diffusion-driven aggregation unlikely, considering the low mobility of oxidized zirconia clusters composed by hundreds and thousands of atoms[34,49];
- the typical time-scale for clusters to reach their steady-state concentration in the PMCS is significantly shorter than the one set by operation of the pulsed source (4 Hz, 250 ms)[50]. Clusters have therefore time to reach their steady-state concentration during a single PMCS pulse.

In the samples with an increasing number of pulses (multiple pulses), the objects with dimension in z-direction (calculated from the histogram of the height in semi-log scale), which differs from the dimensions of primeval incident clusters, have been called islands, according to Jensen.[34] The term island is used regardless whether the structure is resulting from complete coalescence, or juxtaposition in z-direction, as if it is characterized by a spherical, semispherical, or fractal-like shape.

The height, area, and volume distributions are typically log-normal, as it is typical for systems resulting from aggregation processes[51], and they appear Gaussian in a semi-log scale[52]. The distributions have been normalized with respect to the total number of counted particles. A Gaussian fit in the semi-log scale provided the median value of the distribution, while the spread of the distribution was characterized by median absolute deviations (MADs). The error associated to the coverage (not reported in the figures) is affected by the convolution with the AFM tip[40]. It can be considered about 30% for very low coverage and its value decreases with increasing coverage.

In order to investigate the transition from the sub-monolayer to the thin film regime, we have characterized the RMS surface roughness of the samples as a function of coverage, of the number of particles deposited, and eventually of the film thickness. Surface roughness (Rq) is calculated as

$$Rq = \sqrt{\frac{1}{N}\sum_{i,j}(h_{ij} - \bar{h})^2} \qquad (1)$$

where $h_{ij}$ are height values in the topographic map (i, j are the row and column indices) and N is the number of pixels in the map, $\bar{h}$ is the average height ($\bar{h} = \frac{1}{N}\sum_{i,j} h_{ij}$).

We have characterized the evolution of roughness with thickness and with the number of clusters deposited on the substrate, the latter calculated as the total volume of the clusters/islands of the images by the median volume of the multimodal distribution of the first single shot. This is as rough calculation method, and the number of clusters calculated on sample in thin film regime is affected by a further remarkable approximation, since we did not take into consideration the porosity of the film and so we overestimate the number of clusters in the porous matrix. We expect that this error is more pronounced in the ns-ZrOx/He thin films, because preliminary surface analysis measurements suggest a higher porosity in ns-ZrOx/He thin film than in ns-ZrOx/Ar one.

## III. RESULTS AND DISCUSSION

### A. Analysis of the size distribution of the primeval incident clusters

In Fig. 3 we report some representative AFM top-view images of the evolving ns-ZrOx morphology with increasing quantity of deposited clusters of ns-ZrO$_x$/He and ns-ZrO$_x$/Ar films, referred to the centre of the beam. In particular, in Fig. 3(a-b) the topographic images of the first single shot (lowest coverage, θ~2%) are shown; intermediate coverage (θ~50%) is shown in Fig. 3(c-d); the uniform thin film regime, taking place after the 100% coverage limit has been reached, is shown in Fig. 3 (e-f). In all the three coverage conditions, it is qualitatively evident the difference in the topography depending on which carrier gas is used during the deposition: ns-ZrO$_x$/Ar objects appear always higher and larger than the ns-ZrO$_x$/He ones.

The normalized distributions of the heights of the objects analysed in the samples deposited with the first four pulses, are reported in Fig. 4. For each system analysed, the size distributions (height distribution) are broad and multi-modal, and depend on the carrier gas and on the position relative to the beam axis. In both cases it is present a population of very small clusters, with height peaked at 0.4 ± 0.2 nm. The height of these objects is compatible with the deposition of Zr atoms and/or ZrO$_2$ molecules that are present in the cluster beam; the radius of a Zr atom is 0.155 nm[53] and the most compact 3D structure of Zr[54] would be no smaller than 0.57 nm. In order to investigate the real shape of these very small deposited objects we have deconvolved their lateral dimension[55] by estimating the AFM tip radius of 8 nm. We have found an average equivalent width of the particles belonging to the first peaks of 3.4 ± 1 nm. This suggests that the shape of these smallest aggregates is not spherical, but 2D fractal or dendritic-like. It is unlikely these very thin 2D islands are formed in the source chamber or during flight, because they are not energetically favourable. The origin of these 2D islands has to be attributed to the diffusion of Zr atoms on the surface resulting in highly ramified islands[56]. These structures cannot be exactly traced by AFM tip because of its dimension.

The carrier gas strongly affects the size distribution of largest clusters: the size distributions of ns-ZrO$_x$/He and ns-ZrO$_x$/Ar clusters have peaks at 1.9 ± 1 nm and 7.3 ± 4.1 nm, respectively (in agreement with TEM analysis[28], where the average clusters size of ns-ZrO$_x$/Ar is 6.0 ± 1.7 nm). This behaviour is expected because of the different thermodynamic conditions related to the two gases inside the source chamber[57] and it is verified also for cluster-assembled TiO$_x$ films[27]. Selecting the carrier gas therefore allows shifting by a significant amount the median clusters diameter. Inertial effects of clusters in the supersonic beam determine the concentration of larger particles along the beam axis, as proved by the depletion of the large-dimension mode in the case of Ar; in the case of He, depletion is less important, probably because particles in the major mode are already relatively small.

### B. Evolution of the surface coverage

We have defined the coverage as the ratio of the projected area occupied by the clusters on the surface over the area scanned by the AFM. This operative definition of coverage is the same proposed by Jensen for the description of clusters growth,[34] and adopted by others in experimental works[58,59]. It should be noticed that the as-defined surface coverage can be proportional to the deposition time only if a cluster, upon landing on top of a pre-deposited one, quickly diffuses across it, and reaches a free available site on the substrate. However, in the case a diffusion barrier exists at the edge of an island[38,58], the sticking of the new cluster on the preformed island is irreversible, and the evolution of surface coverage no longer follows a proportionality law with respect to the deposition time (or number of clusters deposited). In this case, the evolution of the surface coverage (θ) with time is described by an exponential law[58,60]:

$$\theta = 1 - e^{-\left(\frac{\pi D_m^2}{4}\right) f t} \qquad (2)$$

where $f$ is the average flux (expressed in clusters cm$^{-2}$ s$^{-1}$), $D_m$ is the diameter of the primeval incident clusters, and $t$ is the deposition time. Fig. 5 shows the evolution of the coverage with time (center of the beam). The observed trends can be quantitatively described by equation 5. The diameters of primeval incident clusters extracted by the fit ($D_m$) are 7.9 nm and 2.5 nm for ns-ZrOx/Ar and ns-ZrOx/He clusters, respectively, in very good agreement with the measured effective diameters of the second peaks in the size distributions.

These important results have the following implications:

- The DDA model[38] turns out to accurately describe the evolution of the surface coverage in the case of deposition of preformed clusters from the gas phase, also when a pulsed cluster source is used, and the clusters possess a broad distribution sizes;
- Among the differently sized primeval incident clusters, the larger ones are mainly responsible for the increase of the surface coverage, a conclusion corroborated by fitting eq. 2 to the experimental data;
- The good performance of eq. 2 in describing the evolution of the surface coverage suggests that the diffusion of incoming clusters is strongly inhibited once the substrate is significantly covered by pre-deposited islands;

In order to investigate the growth regime of our interfaces in comparison to the reference cases of diffusion limited, and ballistic deposition (no diffusion),[41] we have characterized the evolution of the vertical width of the growing interfaces, since the latter is not influenced by the limited accuracy of AFM in reproducing the lateral dimensions of nanometer-sized objects.

### C. Evolution of the morphological properties of the islands

The distribution of the diameters of primeval clusters is multi-modal (Fig. 4) and this behavior affects also the morphological properties of the samples at higher coverage, due to different aggregation phenomena.

Fig. 6 shows an example of the evolution of the distribution of the geometrical characteristics of ns-ZrOx/Ar clusters/islands for the first single shot and the subsequent shot (from coverage 3% to 11%). In order to follow the evolution of islands, from the primeval incident clusters through the formation of larger entities, we focused our attention on the highest features formed at a given coverage, represented by the mode of the clusters/islands height distribution with the largest height (not necessarily the most populated one), as indicated by the arrows in Fig. 6. For this reason we will only write about clusters and not atoms/molecules, which anyway characterized the first peak of height distributions shown in Figure 4.

In Fig. 7a-b the evolution of the islands height with surface coverage is shown for ns-$ZrO_x$/He and ns-$ZrO_x$/Ar, respectively. We notice that the dynamics of growth are independent on the region of the film analyzed (center vs. periphery), likely because, as shown in Fig. 6, the same modes are present in the particle size distribution, though with different relative intensities. For this reason, we decided to report hereafter only the results concerning the central part of the beam.

Fig. 7 c-f show the evolution of the projected area and volume of islands with coverage: the maximum coverage investigated is 70%, since for higher coverage the value of the area and volume of objects on the surface increases of several orders of magnitude due to the formation of interconnected structures. These data are affected by the convolution with the AFM tip shape[48]; nevertheless the trends shown are representative.

In the ns-$ZrO_x$/He systems, the average area and volume per island are approximately constant for coverage up to 60%. Also in the ns-$ZrO_x$/Ar systems area and volume are constant for coverage up to coverage 45%, and they grow faster for larger coverage; this evolution reflects the step growth stressed in diameter-z distributions, and which is due to the new-deposited large ns-$ZrO_x$/Ar clusters on preformed surface islands. The area and the volume of ns-$ZrO_x$/Ar islands are systematically larger than those of the ns-$ZrO_x$/He system.

In all the systems analyzed, three regimes can be identified, according to the coverage range considered:

- 0 - 10%. At very low coverage, the coalescence and fast nucleation processes are promoted by the higher diffusivity of the atoms deposited and of the smallest primeval incident clusters[61,34] and by their short time needed to coalesce[61], driven by the minimization of the surface energy[62,63].

- 10 - 70%. For intermediate coverage, the ns-$ZrO_x$/He islands growth in z-direction is frozen, while ns-$ZrO_x$/Ar proceeds stepwise, the second jump in height occurring at about 45% coverage. The data reported in Fig. 7 suggest that the islands growth for ns-$ZrO_x$/He system proceeds via the nucleation of new islands on free surface sites (2D growth); ns-$ZrO_x$/Ar system is characterized by both 2D (before 45% coverage) and 3D growth.

- 70 – 100%. At coverage around 70%, the fast increase in z dimension of islands suggests that a threshold is reached, above which surface diffusion is inhibited because of the presence of pre-deposited clusters (aggregated in islands), acting as pinning centers. This represents the onset of the ballistic deposition regime, where the incoming clusters stick upon landing without significantly diffusing around[41,64,65,66]. Besides the effect of the lateral spatial constraint posed by the pre-deposited particles, the sticking probability likely increases[67], because incident clusters interact now primarily with similar pre-deposited particles. We should take into account that at a surface coverage of 100% the substrate is not necessarily completely covered by clusters/islands. Some voids can be present in the first layer of deposited particles, and masked by the subsequent layers, because of the characteristic growth of surface features typical of the ballistic deposition[41]. According to the definition of coverage adopted here, $\theta \sim 100\%$ means that it is no longer possible to distinguish the substrate in a top-view image, but only the deposited material.

In ns-ZrO$_x$/He systems, the islands growth in z-direction stops very early with coverage, and nucleation events are observed also at high coverage. In Argon systems, few sites (composed by the larger incident clusters) act as nucleation centers for the other smaller and mobile incident clusters promoting the formation of islands from juxtaposition events. In this case, the step growth in z-direction at 45% coverage is probably facilitated by the arrival of new incident clusters on pre-deposited large clusters or islands, which are trapped. In fact, larger clusters mean also a lower dimension[68] and so an open structure of clusters which facilitate the capture. Coverage 70% indicates the starting point of a complete 3D growth for both the systems.

### D. Evolution of clusters and islands density

The data presented in Fig. 7 suggest that the number of new nucleation sites of relatively small islands is higher in ns-ZrO$_x$/He deposition, while the case of ns-ZrO$_x$/Ar islands shape suggests a three-dimensional rather than a two-dimensional growth, whether in both cases, at such a relatively high coverage, this is the result of coalescence or aggregation, it cannot be concluded. Nevertheless we have characterized the evolution of the density of free primeval incident clusters, and of the islands, on the surface (Fig. 8), in order to gain a deeper insight on the growth in the coverage range 10-70%.

The observed qualitative evolution of the surface density of primeval clusters and islands with coverage is similar to the one reported in Ref [58], which refers to the DDA model: for very low coverage the primeval incident clusters density grows leading to a rapid increase of islands density by cluster-cluster aggregation on the surface. This goes on until the islands occupy a small fraction of the surface, roughly 1-10%[34], depending on the incident cluster dimensions. For larger coverages, a competition appears between the nucleation of new islands and growth of existing islands, leading to a slower increase of islands density. Islands density saturates for coverage around 30-50% [34,58,69], when all the incident clusters are captured by previously formed islands, before they can join another cluster and form a new island. At this coverage nucleation becomes negligible. Beyond 30-50% coverage, the linear dimension of the island is comparable to their separation, and islands start merging, which leads to a decrease of the island density[58].

By comparing the cluster-assembled surfaces obtained using He and Ar at the same coverage, we observe a higher island density (or higher density of nucleation sites) for ns-$ZrO_x$/He. The smaller dimensions of He primeval incident clusters provide a larger free surface region for new nucleation events, favored also by the high surface diffusivity of the smaller amorphous ns-$ZrO_x$/He clusters[28].

Some authors report that for small clusters, coalescence is preferred to juxtaposition, and so the occupied area of the new island is smaller than the one occupied by island formed by a juxtaposition process[34]. The faster nucleation events of ns-$ZrO_x$/He primeval incident clusters, compared to ns-$ZrO_x$/Ar, is also demonstrated by the lower coverage needed for the saturation of He-free primeval incident clusters density on the surface ($\theta\sim1\%$) than for Ar ($\theta\sim10\%$) (Fig. 8).

The saturation of island density ($\sigma_{sat}$) is reached around 20-30 % of coverage for Ar; this value is predicted for a growth where only smaller incident clusters can move on the surface and cluster-cluster interactions are prevalently characterized by juxtaposition processes[58,34]. Otherwise, $\sigma_{sat}$ is reached for higher coverage (40-50%) for He system. This behavior can be explained by the possibility that also islands (and not only primeval incident clusters) can move on the surface, by forbidding stationary nucleation sites (which are present in Ar system) for juxtaposition growth. For small He primeval incident clusters, coalescence is preferred to juxtaposition in nucleation processes. In the case of He, the supersonic expansion accelerates ns-$ZrO_x$ clusters towards the substrate at higher velocities compared to Ar[70,8], which facilitates a larger diffusivity of He clusters on the surface[71], and so a higher and faster nucleation events rate.

### E. Interfacial roughening of cluster-assembled nanostructures: from the sub-monolayer to the thin film regime

In order to quantitatively describe the evolution of the interface and to identify a particular growth model, we have characterized the scaling laws, which describe the evolution of roughness with the deposition time. All the objects deposited (atoms or clusters) contribute to the growth of the interface and to the calculation of the RMS roughness. In particular, the evolution of roughness with coverage (Fig. 9) exhibits a nearly linear trend for both systems, but it increases faster in ns-$ZrO_x$/Ar. The slower increase of roughness with coverage for ns-$ZrO_x$/He is due to the smaller incident cluster dimension and islands formed on the surface and to the different growth in the z-direction, consisting in continuous nucleation events, which induces a lateral rather than a vertical growth. This growth dynamics reminds the layer growth mode of atom-assembled thin films[72]. Furthermore, the slope changes approximately around coverage 70%, where the diffusion is inhibited; this point marks the onset of the ballistic deposition regime, as it stems from the study of the evolution of roughness with both the number of deposited particles, and the thickness of the film (Fig. 10).

Fig. 10a reports the increase in surface roughness depending on the number of clusters deposited on the surface (in log-log scale). Since the particle mean flux is approximately constant for He and Ar carrier gas, the number of deposited primeval clusters is approximately proportional to the deposition time; the scaling exponent in the experimental curves of Fig. 10a represents therefore the growth exponent $\beta$ according to the relation $Rq \sim t^\beta$ [73], where t is the thickness of the thin film. We found in the sub-monolayer regime the following values of the growth exponent: $\beta = 0.24 \pm 0.02$ for ns-$ZrO_x$/He, and $\beta = 0.39 \pm 0.02$ for ns-$ZrO_x$/Ar, respectively. The slopes maintain the same values for all the coverage, regardless the diffusion of smallest clusters on the surface, because it does not change the value of the standard deviation of the height on the substrate but only the positioning on the x-y substrate surface.

Our results confirm what is typically obtained in large-scale simulations of the ballistic deposition process; in particular the growth exponent describing the evolution of nanostructures in ballistic deposition model in 2+1 dimensions is 0.24[74]. This value is compatible with $\beta$ of ns-$ZrO_x$/He system, but it does not agree with the one of ns-$ZrO_x$/Ar (that is very close to the value 0.33 of the growth exponent of the 1+1 dimensional system). We have to consider that cluster/cluster interactions in ns-$ZrO_x$/Ar system could be more complicate that in ns-$ZrO_x$/He one, since the former is composed by clusters with a marked two modal size distribution, while primeval He clusters size distribution is more compact. This pronounced difference in size can modulate the sticking probability[67] of primeval clusters on the deposited film. A change in the sticking probability of primeval clusters, principally due to a difference in clusters size, can introduce a change (in particular an increase) in the growth exponent value of ballistic deposition and in the porosity of the film[75,76]. The value $\beta = 0.39$ of the growth exponent of ns-$ZrO_x$/Ar systems agrees well with the value proposed by this modified ballistic deposition model. Remarkably, when one considers the evolution of roughness versus the estimated number of deposited particles, the interfacial roughening process appears to be regulated by the same scaling law across the whole thickness range (from the sub-monolayer regime, up to a coverage of 100%, to the uniform thin film regime). In fact in thin film regime values of the growth exponent are $\beta = 0.20 \pm 0.01$ for ns-$ZrO_x$/He, and $\beta = 0.39 \pm 0.02$ for ns-$ZrO_x$/Ar, respectively. The lower value of $\beta$ in ns-$ZrO_x$/He system compared to the sub-monolayer regime is probably due to the overestimation of the number of clusters in the porous and not compact film. In fact we have not considered the porosity of the film[76], which (also according to the previous considerations regarding the sticking probability) can be estimated higher for ns-$ZrO_x$/He system than for the more compact ns-$ZrO_x$/Ar one, where overhang of clusters is more difficult to produce.

In the uniform thin film regime (Fig. 3e,f), taking place after the 100% coverage limit has been reached, the film thickness t scales proportionally to the deposition time, as long as the deposition rate is constant. The scaling of the surface roughness in this regime can therefore also be described by the roughness vs thickness curve, since $Rq \sim t^\beta$. This is shown in Fig. 10b. The measured growth exponent is $\beta = 0.37 \pm 0.05$ for ns-$ZrO_x$/Ar, and $\beta = 0.32 \pm 0.08$ for ns-$ZrO_x$/He, respectively, again confirming the ballistic deposition regime of the growth. Both these growth exponents are compatible with the ones found by characterizing the evolution of roughness with the number of clusters deposited. The ns-$ZrO_x$/He one is less accurate, but this could also be due to the lacking number of ns-$ZrO_x$/He samples analyzed.

The scaling of the interfacial roughness with incoming clusters number and, later on, with thickness, is therefore independent on the carrier gas and is not influenced by the diffusion of the smallest objects deposited. Nonetheless, the absolute value of Rq, at a given deposition time (which is proportional to the thickness), does depend dramatically on the carrier gas used, as clearly visible in Fig. 3e-f and, quantitatively, in Fig. 10a-b.

## IV. CONCLUSIONS

We have characterized the growth mechanisms affecting the nano- and mesostructure of cluster-assembled films, in particular we studied nanostructured $ZrO_x$ films produced by supersonic cluster beam deposition from sub-monolayer to thin film regime. We have shown that the cluster dimensions prior to deposition (the primeval incident cluster size distribution) affect the growth dynamics, in particular the surface diffusion on the silicon substrate and the nucleation are favored for smaller clusters resulting in a 2D growth, while larger clusters act as static nucleation sites where a 3D growth mode is promoted.

The evolution of the surface coverage with time, and the qualitative trend of primeval incident clusters and islands surface densities, suggest that the DDA model can be used to describe the growth of clusters-assembled film in the sub-monolayer regime, even though the incident flux is pulsed, and the primeval incident clusters are distributed in size.

The growth dynamics in the sub-monolayer regime determines different morphological properties of the cluster-assembled thin film, despite the fact that the evolution of roughness with the number of deposited clusters reproduces exactly the growth exponent of the ballistic deposition in the 2+1 model, across the whole range of coverages, from the sub-monolayer to the thin film regime. Despite qualitatively similar growth mechanisms, the absolute value of the surface roughness of the thin films is strongly influenced by the primeval cluster size. At coverage 70% we identify the onset of the 3D growth. Above this threshold, cluster diffusion is strongly disadvantaged, irrespective of the incident cluster dimension.

Our systematic study gives quantitative information about the fundamental mechanisms of growth of cluster-assembled films, thus providing the ingredients for a deeper theoretical understanding of bottom-up growth processes based on nanoparticle assembling. Our results pave also the way to the quantitative control of the nano- and meso-structure of nanostructured films in view of large scale applications.

**ACKNOWLEDGMENTS**

We thank Paolo Piseri and Cristina Lenardi for useful discussions.

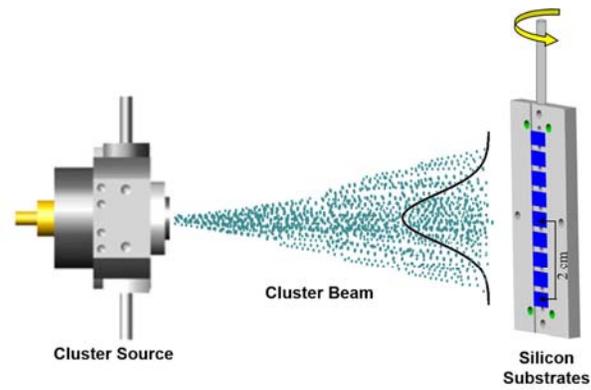

**Figure 1.** Schematic representation of the supersonic cluster beam deposition with a Gaussian intensity profile on a polished silicon substrate, attached to a rotating sample holder. The width of the Gaussian profile at the substrate is 4 cm, approximately

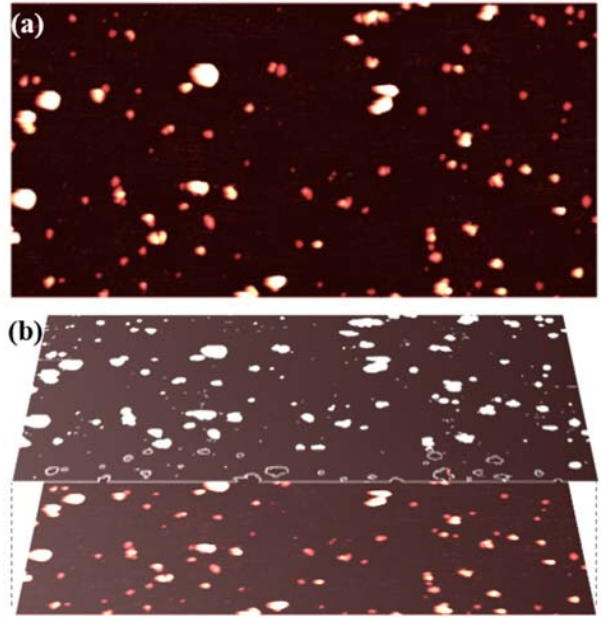

**Figure 2.** (a) AFM top-view topographic map of ns-$ZrO_x$/Ar clusters and islands at low coverage ($\theta \sim 3\%$) (2μm x 1μm, vertical scale is 10nm). In (b - bottom) the same AFM map is shown, in three-dimensional view, with the mask (top) defining the objects to be analysed. The z-threshold has been set at $2\sigma = 0.2$ nm.

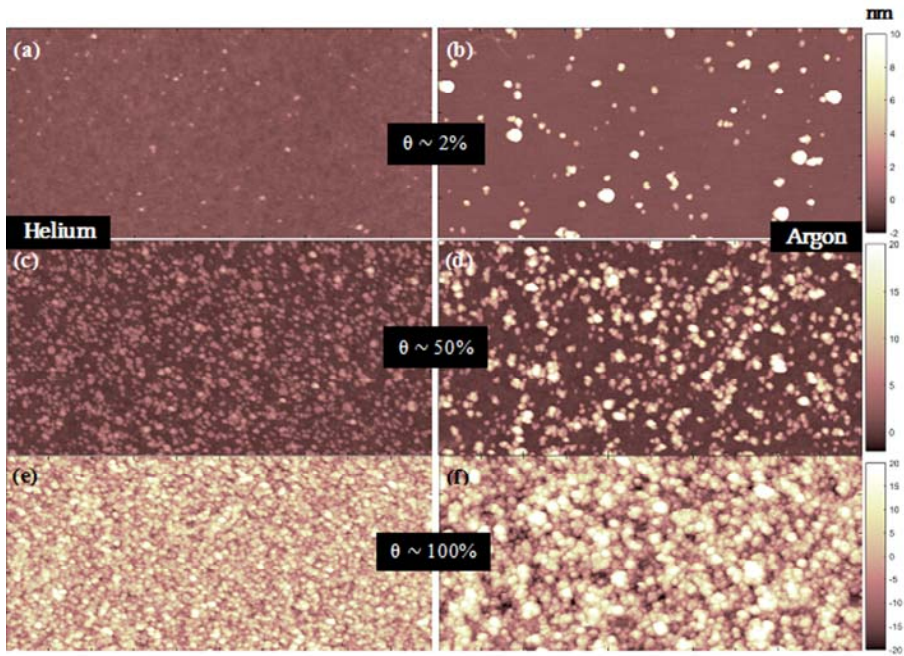

**Figure 3.** (a-b) AFM topographical maps (2μm x 1μm) of ns-ZrO$_x$ thin films for very low coverages (beam center), deposited with Helium and with Argon, respectively; (c-d) ns-ZrO$_x$ sample with coverage θ ~ 50%, (e-f) ns-ZrO$_x$ continuous thin films (thickness about 50 nm).

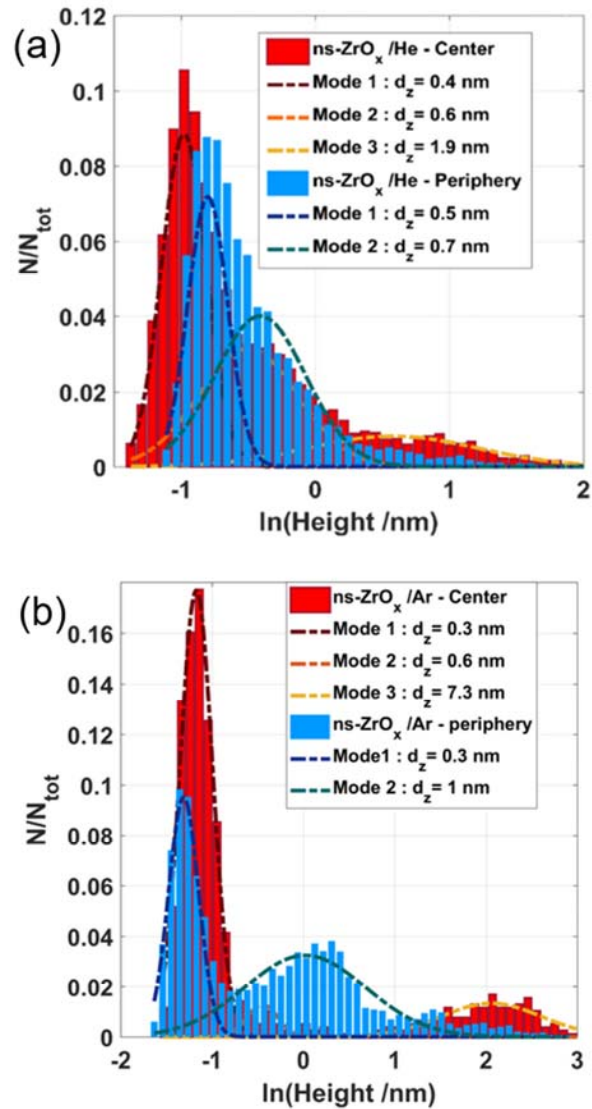

**Figure 4.** Height distributions of the primeval incident clusters. The abscissae represent the logarithms (ln) of the measured heights. (a) ns-ZrOx/He, and (b) ns-ZrOx/Ar, from the center and the periphery of the beam, respectively. The morphologies shown in figure 3a,b have been obtained from cluster beams with the size distributions shown here.

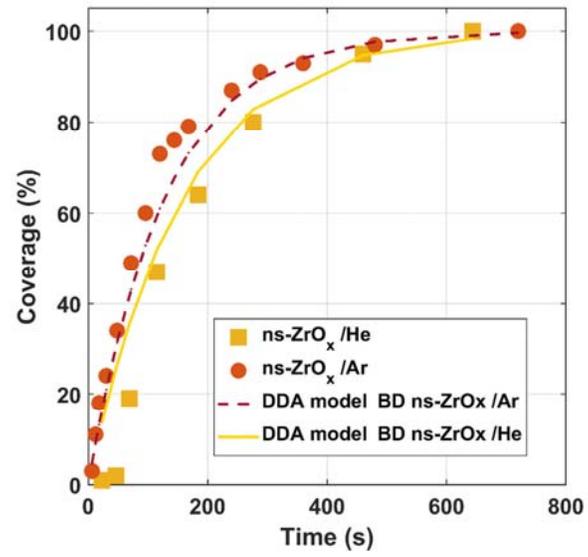

**Figure 5.** Evolution of surface coverage with deposition time.

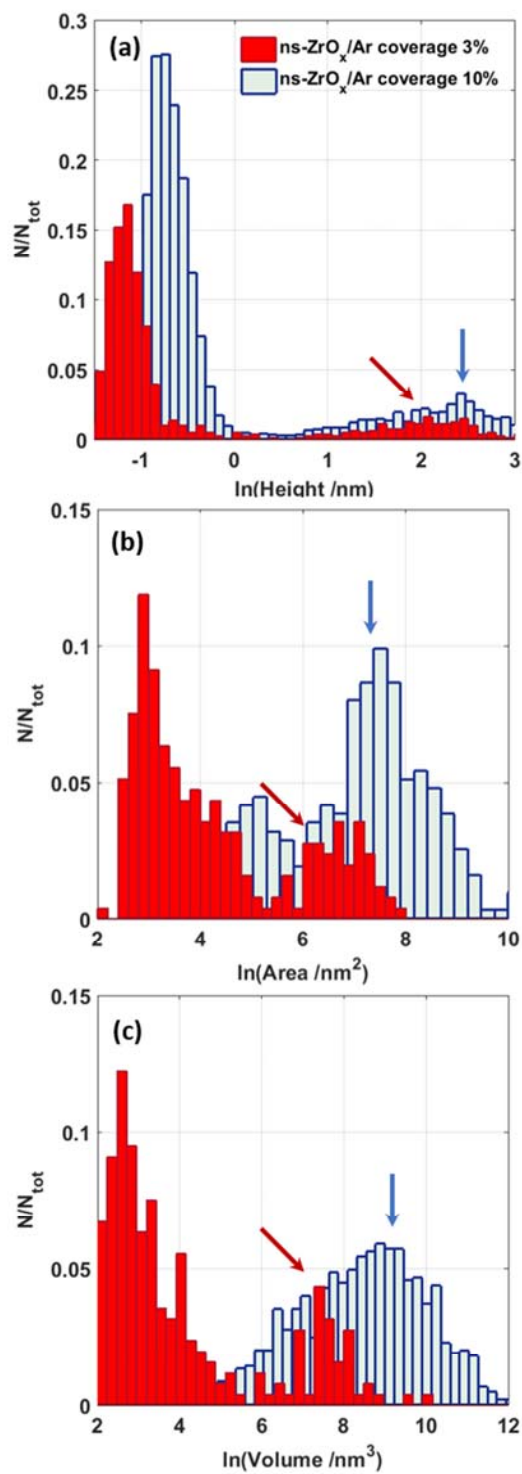

**Figure 6.** Comparison between the distributions of (a) height, (b) area, and (c) volume of ns-ZrOx/Ar clusters/islands of two samples with different coverage (3% and 11%). The abscissae represent the logarithms (ln) of the measured values. The arrows indicate the peaks of the largest objects, whose mean values are used to describe the evolution of the geometrical properties with coverage.

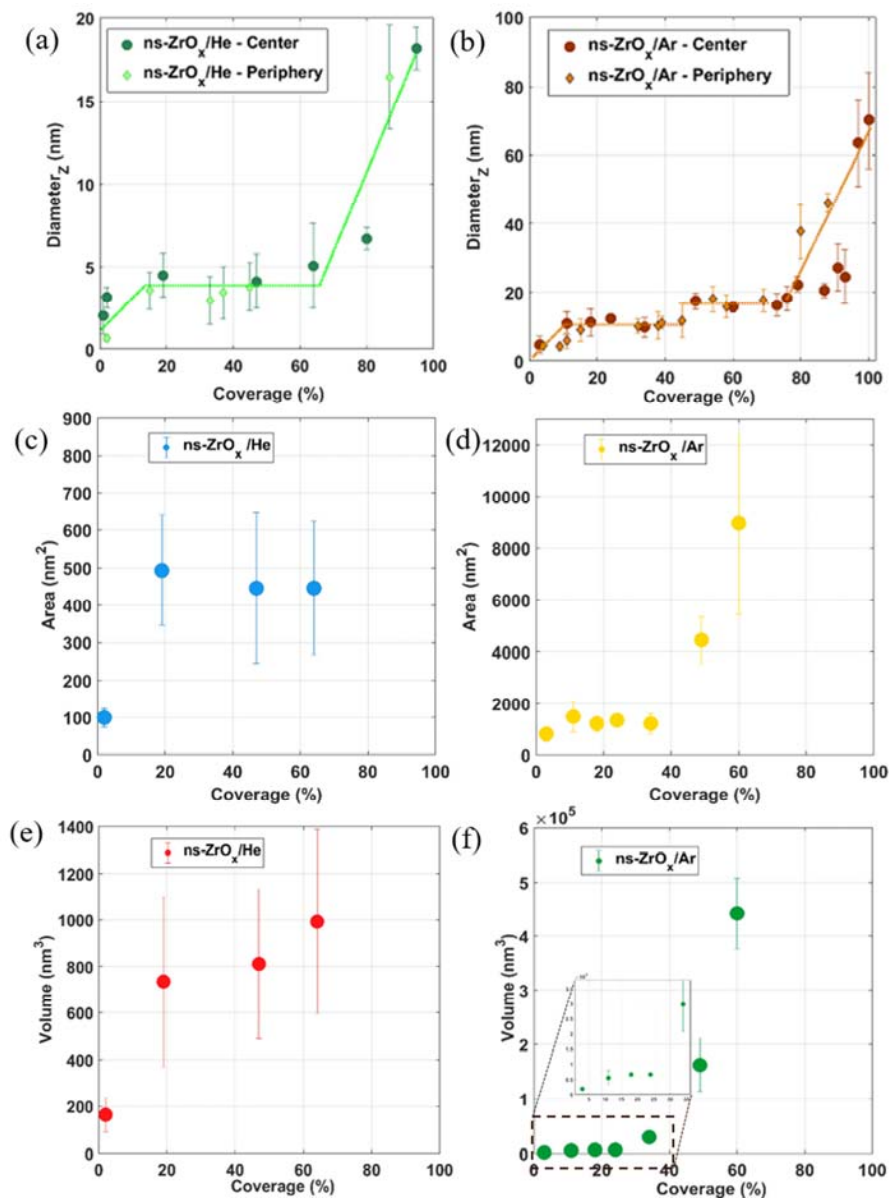

**Figure 7.** Evolution of different properties of zirconia islands with coverage, calculated by the AFM topographical maps, for ns-ZrOx/He and ns-ZrOx/Ar. (a-b) z-diameter on different regions of the film (center or periphery); (c-d) projected area; (e-f) volume.

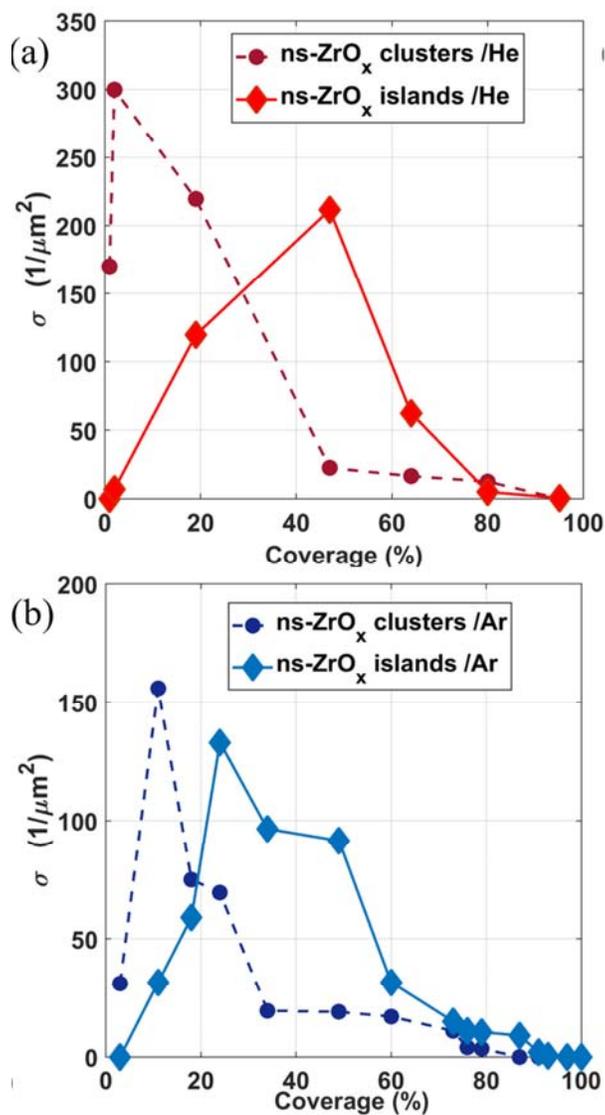

**Figure 8.** Evolution of the primeval incident (a) ns-ZrOx /He, and (b) and ns-ZrOx /Ar clusters and islands densities (σ), as a function of the surface coverage, in the central region of the beam.

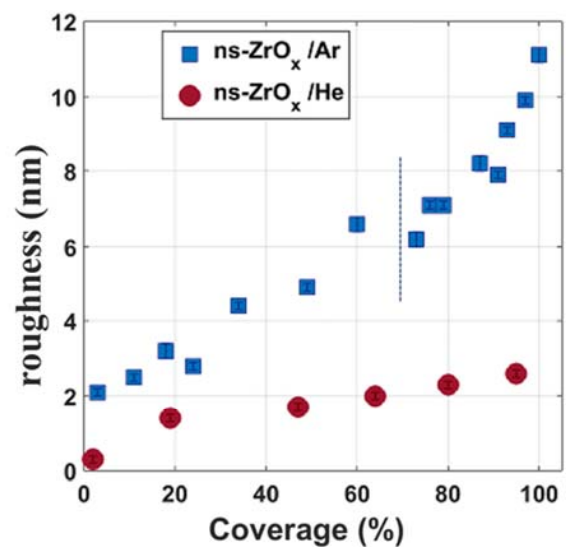

**Figure 9.** Evolution of the rms surface roughness Rq with coverage in the sub-monolayer regime.

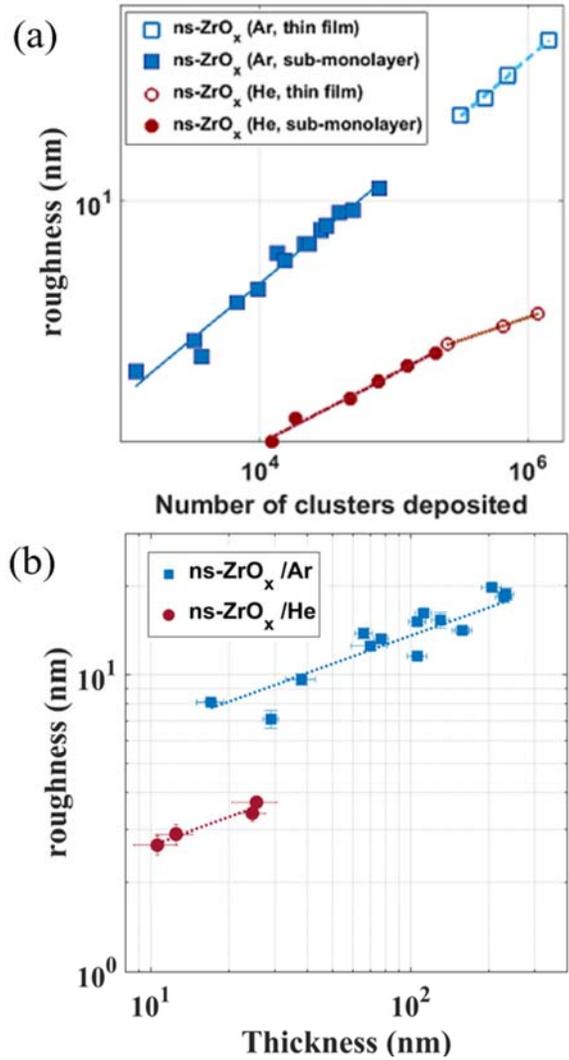

**Figure 10.** (a) Evolution of the rms surface roughness Rq with the estimated number of clusters deposited (in log-log scale), from the sub-monolayer to the continuous thin film regime. (b) Scaling of the surface roughness with film thickness, for helium and argon as carrier gas. The linear regressions of the experimental curve Rq ~ $h^\beta$ is also shown.